\documentclass[aps,showpacs,prb,twocolumn,superscriptaddress]{revtex4}
\usepackage{times,xspace}
\usepackage{amsbsy,amssymb,amsmath,bm}
\usepackage{graphicx,color,epsfig,rotate}
\usepackage{fancyheadings}

\begin{document}
\title{ELASTIC DEFORMATION OF POLYCRYSTALS}
\author{Rajeev Ahluwalia} 
\author{Turab Lookman}
\author{Avadh Saxena}
\affiliation{Theoretical Division, Los
Alamos
National Laboratory,
Los Alamos, New Mexico, 87545}
\date{\today}

\begin{abstract}
We propose a framework to model elastic properties of
polycrystals by coupling crystal orientational degrees
of freedom with elastic strains. Our model encodes crystal
symmetries  and takes into account explicitly
the strain compatibility induced long-range interaction between grains. 
The  coupling of crystal orientation and elastic interactions 
allows for the rotation of  individual grains by an external load.
We  apply the model to simulate  uniaxial tensile loading of a
$2D$ polycrystal within linear elasticity and a system 
with elastic anharmonicities that describe
structural phase transformations.
We investigate the constitutive response of the polycrystal
and compare it to that of single crystals with  
crystallographic orientations that form the polycrystal.
\end{abstract}
\maketitle
 A study
of the mechanical properties of polycrystals is important 
as  most technologically
important materials exist in a polycrystalline state. 
A polycrystal is an aggregate of grains that have different
crystallographic orientations. The properties of a polycrystal
depend on its texture (distribution of crystallographic
orientations). It is desirable to understand how the strains 
are distributed due to an applied external load  and how 
this  influences the average elastic moduli of
the polycrystal \cite{avell}. An important feature of some polycrystalline
metallic alloys and ceramics is the presence of domain walls
within the grains due to an underlying martensitic
 transition. This microstructure influences the
response of the material to an external load. For example, in
martensites, the deformation is accompanied by the motion of the
 domain walls. 
 Texture evolution caused by rotation of grains
under the application of an external load is another important
factor that influences the mechanical behavior of a polycrystal.
Polycrystalline specimens exhibit significant grain rotation in
the plastic regime to accommodate crystallographic slip
\cite{poulsen}. Grain rotations up to $\sim 1^{o}$ have also been observed in 
atomistic simulations 
\cite{schiotz,swy} 
of nanocrystals, even at low strains of $\sim 4\%$.

The problem of finding the effective properties of
polycrystals has been studied by analytical methods
\cite{avell,bhatt1}. However, 
the complex geometry of a polycrystal or the long-range elastic
interactions between the
grains are often not accounted for in these methods.
In fact, these approaches are only able to give bounds on
the effective properties. Recently, the mechanical deformation of
polycrystals has been studied by atomistic
simulations \cite{schiotz,swy} which have been limited
to nano-sized grains. Simulating bulk systems with atomistic
simulations requires enormous computational power and hence continuum
simulations that can cover a range of intermediate
length scales are essential for describing the 
microstructure at the sub-micron scale. Several
phase-field models have been proposed to model grain growth
phenomena \cite{elder1,chen,warren}. Although these models
correctly describe the grain morphologies and the domain growth
laws, the issues of elasticity and material specific 
crystal symmetries are usually  not addressed. Recently Elder et al. \cite{elder2}
studied elastic and plastic effects using a model
formulated to describe pattern selection. 
The model, in terms of particle density fields, 
is specific  only to certain  symmetries that are selected
by appropriate choice of wavelengths in the free energy, which 
does not contain  experimentally
measured quantities such as elastic constants. In this Letter,
we propose a polycrystal model based on continuum elasticity that
can be applied to any crystal symmetry and has the
appropriate single-crystal elastic constants as input parameters.
In this model, elastic strains are coupled to a phase field model
through an orientation field that is determined from a
multi-component order parameter describing the crystal
orientations. Due to this coupling, the strains in each grain as well as
the grain orientations can change under an
external load. This experimentally relevant feature is
not accounted for in models that consider static grains created by Voronoi
construction \cite{khacha1}. In the present work, we
determine the mechanical properties of linear elastic materials and  
those described by nonlinear elasticity, such as martensites.

The free-energy functional is written as
$F=F_{grain} + F_{elastic} +F_{load},$
where $F_{grain}$ is the  free energy density due to the
orientational degrees of freedom of the polycrystal, $F_{elastic}$
represents the elastic free energy and $F_{load}$ is the free
energy contribution due to an external applied load. The
polycrystalline system is described by a set of $Q$ non-conserved
order parameters \cite{chen} $(\eta_{1},\eta_{2},...,\eta_{Q})$.
A given grain orientation corresponds to 
one of the $Q$ order
parameters being positive nonzero while the rest are zero.
The free energy $F_{grain}$ is
given by
\begin{eqnarray}
F_{grain}&=&\int d\vec{r}\bigg\{\sum^{Q}_{i=1}\bigg[{{a_1}\over{2}}{\eta_i}^2
+{{a_2}\over{3}}{\eta_i}^3 +{{a_3}\over{4}}{\eta_i}^4\bigg]\nonumber\\
&+&
{{a_4}\over{2}} \sum^{Q}_{i=1} \sum^{Q}_{j>i}{\eta_i}^2 {\eta_j}^2
+\sum^{Q}_{i=1} {{K}\over{2}}(\nabla \eta_i)^{2}\bigg\}.
\end{eqnarray}
For $a_1,a_2 < 0$ and $a_3,a_4 > 0$, the first two terms in
equation (1) describe a potential with $Q$ degenerate minima
corresponding to $Q$ grain orientations. The gradient energy $(K >
0)$ represents the energy cost of creating a grain boundary.
 It is also possible to associate an orientational field 
$\theta(\vec{\eta},\vec{r})$, where 
\begin{equation}
\theta(\vec{\eta},\vec{r})={{\theta_m}\over{Q-1}}
\bigg[{{\sum^{Q}_{i=1} i\eta_i}\over {\sum^{Q}_{i=1}
\eta_i}}-1\bigg].
\end{equation}
Thus, there are $Q$ orientations between $0$ and $\theta_{m}$. 
For the elastic free energy the linearized strain tensor
in a global reference frame is
 defined by $\epsilon_{ij}=(u_{i,j}+u_{j,i})/2$ $(i=1,2$: $j=1,2)$, 
where $u_i$ represents $i{\mbox{th}}$ component of the
displacement vector and $u_{i,j}$ is its $j{\mbox{th}}$ 
displacement gradient.  For illustration, we consider a $2D$ lattice 
with square symmetry and use
the symmetry-adapted linear combinations of the strain tensor
defined as \cite{krum} $\epsilon_1=(\epsilon_{xx}+\epsilon_{yy})/\sqrt2$,
$\epsilon_2=(\epsilon_{xx}-\epsilon_{yy})/\sqrt2$ and
$\epsilon_3=\epsilon_{xy}$. 
To generalize this theory
for the case of a polycrystal, the strain tensor in a
rotated frame is calculated as $R(\theta(\vec{\eta}))\epsilon
R^{T}(\theta(\vec{\eta}))$, where $R(\theta(\vec{\eta}))$ is a
rotation matrix. Using this transformation, the elastic free energy 
in a global frame of reference is 
$F_{elastic}=\int d\vec{r}\{{{A_1}\over{2}}{e_1}^2 +{{A_2}\over{2}}{e_2}^2
+{{A_3}\over{2}}{e_3}^2+{f_{nl}}({e_1},{e_2},{e_3})
 +{{K_2}\over{2}}{(\nabla e_2)}^2
+{{K_3}\over{2}}{(\nabla e_3})^2\}$
where  $e_1$, $e_2$, $e_3$
are defined as $e_1=\epsilon_1$,
$e_2={\epsilon_2}\cos[2\theta(\vec{\eta})]+\sqrt{2}{\epsilon_3}
\sin[2\theta(\vec{\eta})]$
and
$e_3=-(1/\sqrt{2}){\epsilon_2}\sin[2\theta(\vec{\eta})]+{\epsilon_3}
\cos[2\theta(\vec{\eta})]$. The orientation field $\theta(\vec{\eta})$ is 
determined from the minima of free energy in (1) using 
(2). 
Here $A_1=C_{11}+C_{12}$, $A_2=C_{11}-C_{12}$ and $A_3=4C_{44}$,
where $C_{11}$, $C_{12}$  and $C_{44}$ are the elastic constants
for a crystal with square symmetry. $K_2$ and $K_3$ are the
appropriate gradient coefficients that in principle can be
obtained from experimentally measured phonon dispersion data. The term
$f_{nl}({e_1},{e_2},{e_3})$ represents the nonlinear part of the
elastic free energy and is crucial in describing structural
phase transitions. 

In this work, we are interested in simulating a
uniaxial loading experiment. If we choose the $x$ axis to be the
loading direction,  the free energy contribution due to the
external load is 
$F_{load}=-\int d\vec{r}\sigma \epsilon_{xx} 
=-\int d\vec{r}{{\sigma}\over{\sqrt{2}}} (\epsilon_{1}+\epsilon_{2})
=-\int d\vec{r}{{\sigma}\over{\sqrt{2}}}
(e_{1}+{e_2}\cos[2\theta(\vec{\eta})]
-\sqrt{2}{e_{3}}\sin[2\theta(\vec{\eta})]).$
The strains $\epsilon_{1}$
, $\epsilon_{2}$
and $\epsilon_{3}$ are not independent but satisfy a compatibility 
relationship 
\cite{love}: 
${{\nabla}^2}{\epsilon_1}-({{\partial^2}\over{\partial x^2}}
-{{\partial^2}\over{\partial y^2}}){\epsilon_2}-\sqrt{8}
{{\partial^2}\over{ {\partial x}{\partial y} }}{\epsilon_3}=0.$
Using a method introduced earlier for single
crystal martensitic transformations \cite{lookman},
the strain $e_1$ may be eliminated using
compatibility, to express the  effective free energy
$F_{eff}=F_{elastic}+F_{load}$  as
\begin{eqnarray}
F_{eff}&=&{{A_1}\over{2}}\int d\vec{k}\bigg[ 
{C_2}^2(\vec{k})
|\Gamma_2(\vec{k})|^{2} +
{C_3}^2(\vec{k})|\Gamma_3(\vec{k})|^{2}\nonumber\\ &+&
{C_2(\vec{k})}{C_3(\vec{k})} [\Gamma_3(\vec{k})
\Gamma_2(-\vec{k})+ \Gamma_3(-\vec{k})
\Gamma_2(\vec{k})]\bigg] \nonumber \\ &+&\int
d\vec{r}\bigg[{{A_2}\over{2}}{e_2}^2
+{{A_3}\over{2}}{e_3}^2+{f_{nl}}({e_2},{e_3})\nonumber\\
&+&{{K_2}\over{2}}{(\nabla e_2)}^2 +{{K_3}\over{2}}{(\nabla
e_3})^2\nonumber\\ &-&{ {\sigma}\over{\sqrt{2}} }
\bigg({e_2}\cos[2\theta(\vec{\eta})]
-\sqrt{2}{e_3}\sin[2\theta(\vec{\eta})]\bigg)\bigg] .
\end{eqnarray}
where $\Gamma_2(\vec{k})$, $\Gamma_3(\vec{k})$  represent  Fourier transforms of
${e_2}\cos[2\theta(\vec{\eta})]-\sqrt{2}{e_3}\sin[2\theta(\vec{\eta})]$
and 
${e_2}(\sin[2\theta(\vec{\eta})]/\sqrt{2})+{e_3}\cos[2\theta(\vec{\eta})]$ respectively,
$C_2(\vec{k})=({k_{x}}^2-{k_y}^2)
/({k_{x}}^2+{k_y}^2)$ and
$C_3(\vec{k})=\sqrt{8}{k_{x}}{k_y}
/({k_{x}}^2+{k_y}^2)$.
The long-range terms ensure
that compatibility is satisfied within the grains as well as at the
grain boundaries.

The dynamics of the grains is given by $Q$
 equations  
$${{\partial \eta_i}\over{\partial t}}=-
{\gamma_{\eta}}{{\delta F}\over{\delta \eta_i}},  \eqno(4)$$
where $\gamma_{\eta}$ is a dissipation coefficient and $i=1,...,Q$ 
correspond to $Q$ grain orientations.
The corresponding overdamped dynamics for the strains is 
$${{ \partial e_2} \over {\partial t} }
=-{\gamma_2}\left[{{\delta F}\over{\delta e_2} }
\right], 
{{ \partial e_3} \over {\partial t} }
=-{\gamma_3}\left[{{\delta F}\over{\delta e_3} }
\right], \eqno(5)$$
where $\gamma_{2}$ and $\gamma_{3}$ are the appropriate dissipation
coefficients for the strains and
$F=F_{grain}+F_{eff}$
is the total free energy of the system. 


The mechanical properties of many
materials are well described by the harmonic approximation
for which the nonlinear term
${f_{nl}}({e_1},{e_2},{e_3})=0$. 
For a homogeneous single crystal $e_2$, $e_3$ and 
$\vec{\eta}$ are constant and for Cu     
$A_1=289.8$ GPa, $A_2=47.0$ GPa and $A_3=301.6$ GPa (
$C_{11}=168.4$ GPa, $C_{12}=121.4$ GPa,
$C_{44}=75.4$ GPa). For the 
parameters in $F_{grain}$ we choose  $a_1=a_2=-A_2$, $a_3=A_2$, 
$a_4=2A_2$, $Q=5$
and ${\theta_m}=45^o$.
We choose the 
gradient coefficients in terms of an arbitrary length scale
$\delta$ so that $K=K_2=K_3={A_2}{\delta}^2$ and lengths are scaled by  
$\vec{r}=\delta\vec{\zeta}$.
The free energy in
(1) then has five degenerate minima defined by
$\theta_0(\vec{\eta})={0^o}, {11.25^o},{22.5^o},{33.75^o},45^o$, corresponding 
to five different grain orientations.  

To study the
polycrystal, we first generate an initial polycrystalline
configuration by solving (4) and (5) with 
$\sigma=0$, using random initial conditions.
For all simulations in this paper, we assume $\gamma_{\eta}=\gamma_{1}=\gamma_{2}=\gamma$ and 
use rescaled time $t^*=t{|A_{2}|}\gamma$. 
For $\sigma=0$, the
elastic effects do not influence the grain growth as all the
strains vanish. Grains with orientations
$\theta_0(\vec{\eta})={0^o}, {11.25^o},{22.5^o},{33.75^o},45^o$
nucleate and coarsen. We arrest the system in a given 
configuration by suddenly changing the value of the parameter
$a_1$ from $-A_2$ to $-16A_2$. This 
increases the free-energy barriers between the crystalline states
and the growth stops. With the arrested polycrystal configuration
as the initial condition, we simulate a quasi-static uniaxial
tensile loading using  (4) and (5).  
The stress $\sigma$ is
varied in steps of $0.06$ GPa and we let the strains relax after each
change for $t^*=25$ steps. Figure 1(a) shows the spatial distribution of the
polycrystal orientation ${\theta_{\sigma}}(\vec{r})$ at a loading
of $\sigma=2.35$ GPa for a system of size 
$128\delta\times128\delta$. We note that there is no
significant motion of the grain boundaries from the initial
arrested configuration to the configuration depicted in Fig.
1(a) (the individual grains have rotated by a small
amount $(\sim 0.01^{o})$, consistent with the coupling between stress and the
orientation). 
 In Fig. 1(b), we show the corresponding distribution of
uniaxial strain $\epsilon_{xx}$. 
The strain distribution is 
anisotropic as $\epsilon_{xx}$ in a grain
depends on the orientation.

In Fig. 2, we show the variation of
the average strain $\langle\epsilon_{xx}\rangle$ with the load $\sigma$.
For comparison, we also plot the
analogous single crystal curves with 
crystallographic orientations that constitute the polycrystal
configuration in Fig. 1 (single crystal simulations were performed 
  using  only one orientation but with 
identical free energy parameters and loading rate as the polycrystal). 
 The Young's modulus of the simulated
polycrystal 
was $\sim 126$ GPa. 
This is in the range of experimentally measured values
of $124$ GPa \cite{ym} and $129.8$ GPa \cite{hertz} quoted for bulk 
polycrystalline Cu.
The result is not sensitive to the choice of parameters
for the polycrystal phase field model, at least in the linear elastic
regime. 



Another important class of materials that can be studied using this approach 
are martensites that undergo a displacive structural phase transformation.
The transformed phase is characterized by
a complex arrangement of crystallographic variants known as twins. 
We consider the case of a $2D$ square to
rectangle transition for which the deviatoric strain
$e_2$ is the appropriate order parameter. For the high temperature
square phase $e_{2}=0$ and for the low temperature martensitic
phase $e_2=\pm e_0$, corresponding to the two rectangular
variants. This system exhibits the so
called shape-memory effect  
which is governed by the motion of  martensitic domains.
The microstructure 
depends on the underlying crystal symmetry and hence the 
displacements and domain wall orientations 
of the atoms in each grain depend on the grain
orientation. Thus the shape memory effect will be
influenced by the texture and hence it is important  
to compare the mechanical response of single and polycrystal martensites. 

The anharmonic contribution to the elastic free energy 
is given by 
$f_{nl}={{\alpha}\over{4}}{e_2}^{4}+{{\beta}\over{6}}{e_2}^{6}$
and describes a first
order transition for $\alpha < 0$. We choose 
$A_1=140$ GPa, $A_3=280$ GPa, $\alpha=-1.7\times{10^4}$ GPa and
$\beta=3\times{10^7}$ GPa, parameters that correspond to 
FePd \cite{kartha}.  
The constant $A_2$ depends on the
temperature and we choose $A_2=-3$ GPa, 
a temperature in the martensitic phase.  
The measured gradient coefficient $K_2/{{a_0}^2}=25$ GPa, where $a_0$
is the lattice spacing of the crystal and assume $K_3=0$ since the
deviatoric strain is the dominant mode of deformation. The
parameters for $F_{grain}$ are $a_1=a_2=-|A_2|$,
$a_3=|A_2|$, $a_4=2|A_2|$ and the grain boundary coefficient is
chosen in terms of the lattice spacing $a_0$ to be 
$K/{{a_0}^2}={10^4}{|A_2|}$. The space variable is rescaled by
introducing a dimensionless length scale $\vec{\zeta}$ so that
$\vec{r}=(100{a_0})\vec{\zeta}$. 
The maximum orientation is
chosen to be $\theta_m=30^o$ so for $Q=5$, the allowed
orientations are $0^o$, $7.5^o$, $15^o$, $22.5^o$ and $30^o$. Employing
the above set of parameters, we simulate the martensitic domain
structures using (4) and (5). With same
procedure as for the linear elastic case, a stable polycrystal
configuration is obtained. When the applied stress $\sigma=0$, the
parameter $A_2=-3$ GPa ensures that the system is well in the
martensite phase and domains of the two rectangular
variants (twins) are formed.
After obtaining a stable martensitic polycrystal, the loading process 
is simulated by quasi-statically varying the stress in steps of $0.38$ GPa upto
a maximum stress $15$ GPa 
(after each stress change, the system is allowed to relax for 
$t^{*}=25$ steps).
The system is then unloaded by decreasing the stress
to zero at the same rate.

Figure 3 shows the evolution of variants and the grains at different 
stress levels during the loading-unloading process for a system of size 
$12800{a_{0}}\times 12800{a_{0}}$.  The left column shows the
distribution of $\epsilon_2(\vec{r})$ (deviatoric strain in a global frame) and
the right column shows the corresponding distribution of the orientations 
$\theta(\vec{r})$. It is clear from Fig. 3 that the domain wall orientations
depend on the orientation of a grain. On loading, the simulated polycrystal 
starts to detwin (favored variants grow at the expense of unfavored ones).
However, even at the maximum load of $\sigma=15$ GPa, some unfavorable variants
persist. On complete unloading, a domain structure is 
nucleated again due to inhomogeneities in the polycrystal. However, this
domain structure is not the same as that before the loading indicating an 
underlying hysteresis. The orientation distribution is also influenced by the
external load, as can be seen in the right column of Fig. 3. 
The grains with large misorientation with the loading axis rotate significantly
$(\sim 10^{o})$ while the grains having lower orientation do not rotate as much.
The mechanism of this rotation is the desire of the system to maximize strain
in the loading direction so as to minimize the elastic free energy. At high
stress, some grain boundaries start moving to accommodate the applied 
stress, as is clear from the orientation distribution at $\sigma=15$ GPa.

The stress-strain curve corresponding to Fig. 3 is
shown in Fig. 4. Also shown are single crystal curves for all 
five orientations that constitute the polycrystal of Fig. 3.  We observe that
the hysteresis for the polycrystal case is much smaller 
than that for a single crystal oriented along the loading axis. 
Our findings regarding the hysteresis are 
consistent with the fact that polycrystals have poor 
shape memory properties compared to single crystals \cite{bhatt1}. The 
simulations also indicate that grain rotations will influence the 
mechanical properties of shape memory alloys.  
Recently, {\it in situ} measurements of texture evolution during compression 
experiments on Ni-Ti 
shape memory alloys \cite{vaidya} have been reported. However, these 
experiments cannot predict whether the changes in texture are due to detwinning
or rotation of grains. Our simulations show that both these processes can
contribute to texture evolution. 

In summary, we have proposed a  framework to study the
mechanical properties of polycrystals in which the long-range 
elastic interaction between grains and the
connectivity of the microstructure is taken into account.
The approach can be 
extended to any crystal symmetry or loading (e.g., shear) and does 
not require any a priori assumption of grain shapes or the microstructure. 
 An important
feature of our work is the coupling between the grain
orientation and elasticity. 
We have applied the model to study
mechanical properties of linear elastic and martensitic materials. 
For the linear elastic case
the observed grain rotations  are small 
$(\sim 0.01^o)$ and hence do not influence the mechanical properties.
In contrast, 
 the martensitic case shows 
significant grain rotation $(\sim 10^{o})$ due to accommodation
of the transformation strain. This behavior is sensitive to the choice
of parameters of the polycrystal model (energy barriers between grains) 
and therefore determination of these parameters from experiment or
atomistic simulations will allow accurate prediction of mechanical properties.

We thank Kai Kadau and R.C. Albers for discussions.  This work was supported by the 
U.S. Dapartment of Energy. 

\newpage
Figure Captions:

Figure 1: Spatial distribution of orientation angle $\theta(\vec{r})$
(snapshot (a))
and uniaxial strain
$\epsilon_{xx}(\vec{r})$ (snapshot (b)) for stress $\sigma=2.35$ GPa.

Figure 2: Average uniaxial strain $\langle\epsilon_{xx}\rangle$ for the linear
elastic case as a
function of the load $\sigma$. The curves correspond to a polycrystal
($\circ$) and single crystals with $\theta_0=0^{o}$
($\times$), $\theta_0=11.25^{o}$ ($+$),
$\theta_0=22.5^{o}$($*$),
$\theta_0=33.75^{o}$($\square$)
and $\theta_0=45^{o}$ ($\diamond$).

Figure 3: Spatial distribution of the deviatoric strain in a global frame,
$\epsilon_2(\vec{r})$, (snapshots (a),(c),(e) and (g)) and orientation
angle $\theta(\vec{r})$ (snapshots (b),(d),(f) and (h)). The corresponding
stress levels are $\sigma=0$ ((a) and (b)),
$\sigma=2.69$ GPa ((c) and (d)),
$\sigma=15$ GPa ((e) and (f)) and
$\sigma=0$ (after unloading) ((g) and (h)).

Figure 4: Average uniaxial strain $\langle\epsilon_{xx}\rangle$ for the
martensite as a
function of the load $\sigma$. The curves correspond to a polycrystal
($\circ$) and single crystals with $\theta_0=0^{o}$
($\times$), $\theta_0=7.5^{o}$($+$),
$\theta_0=15^{o}$($*$),
$\theta_0=22.5^{o}$($\square$)
and $\theta_0=30^{o}$($\diamond$).
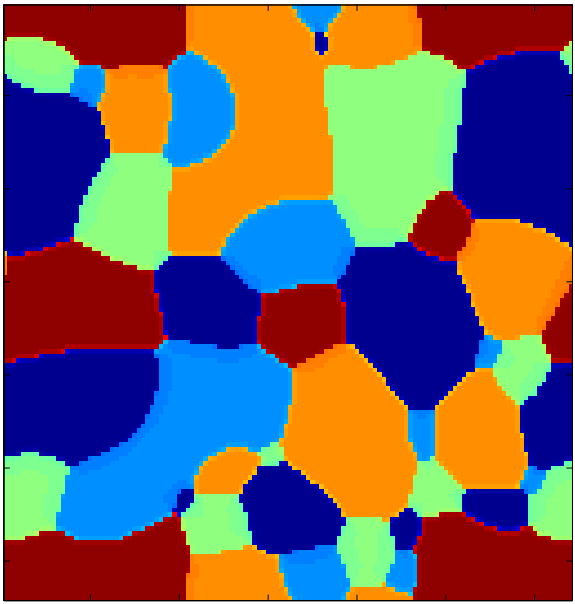
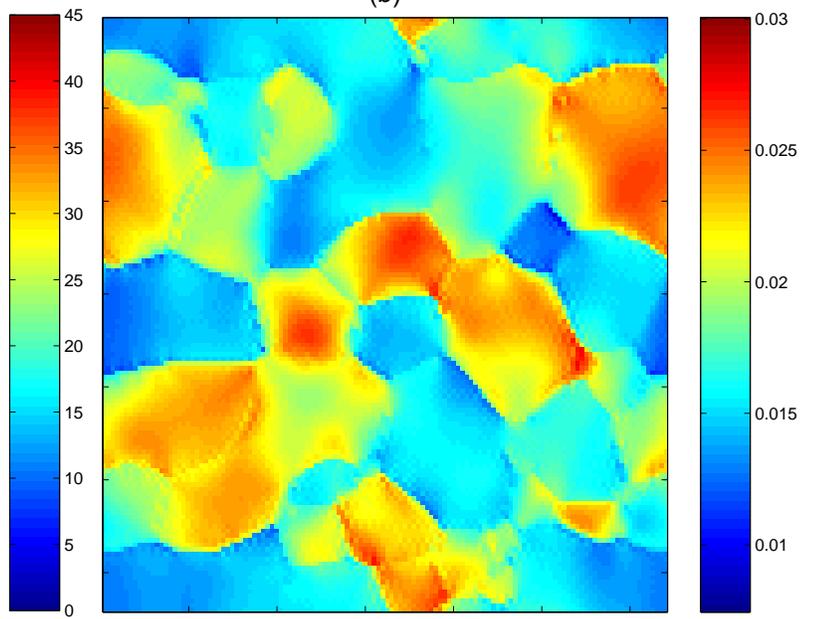
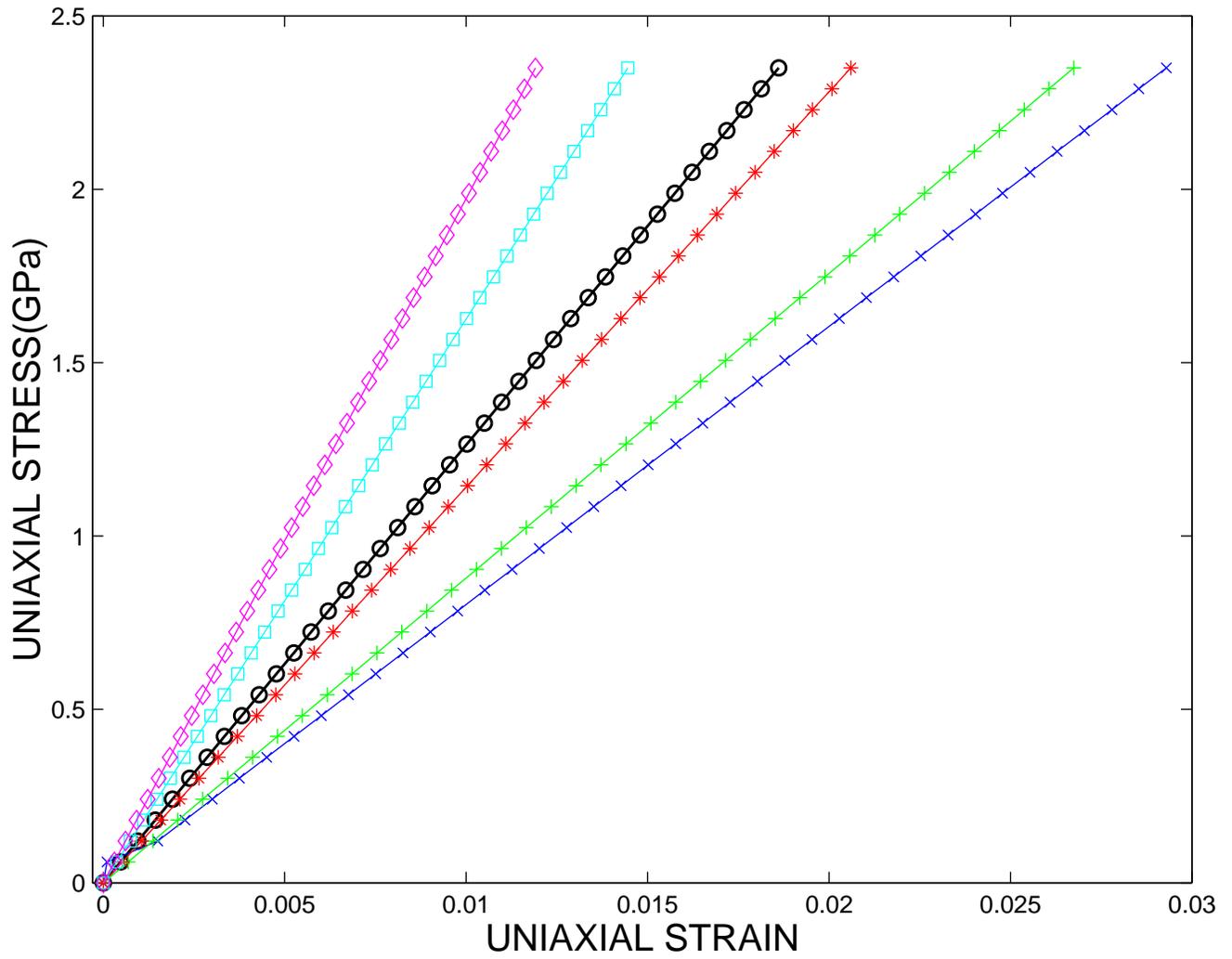
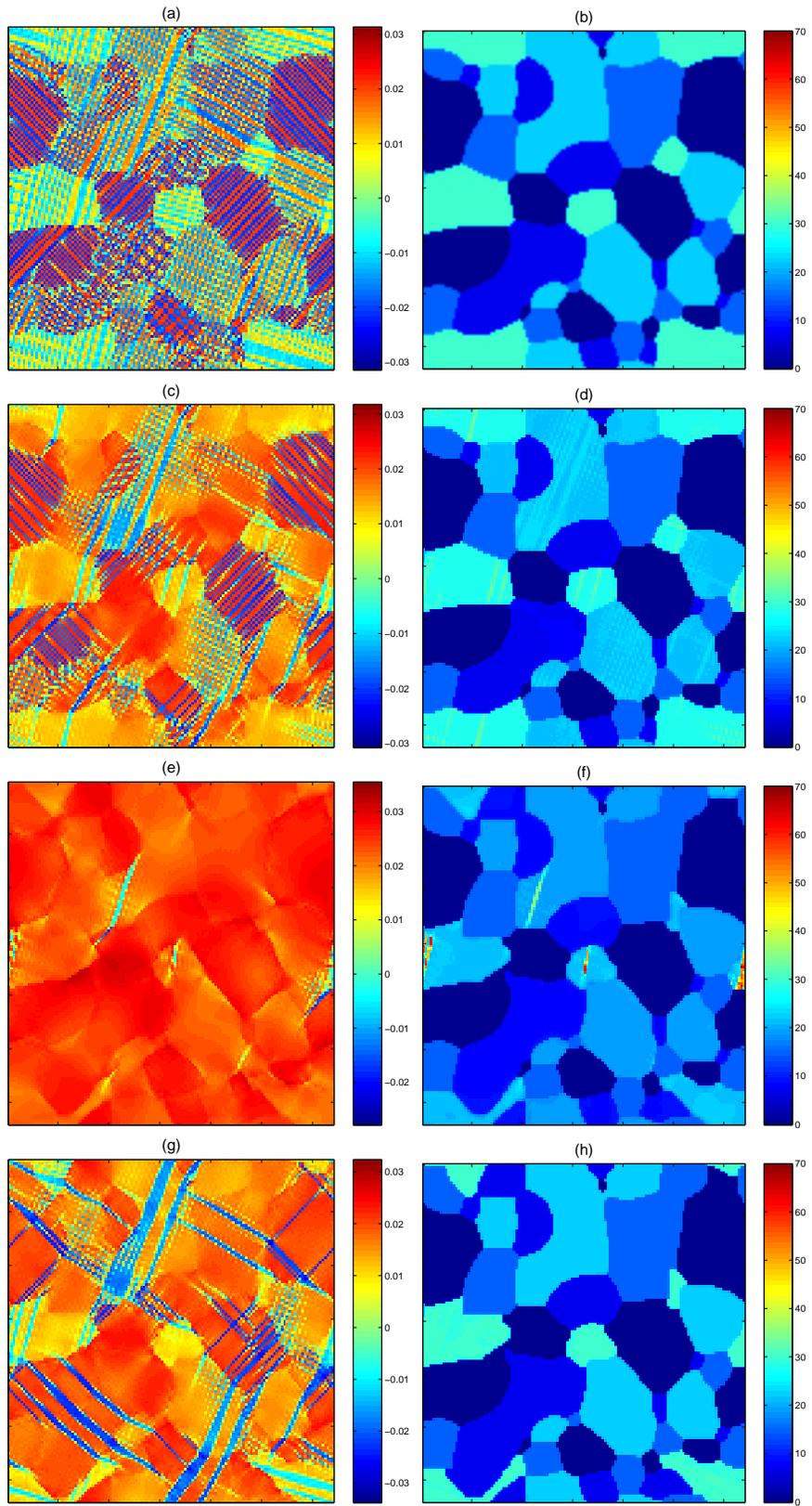
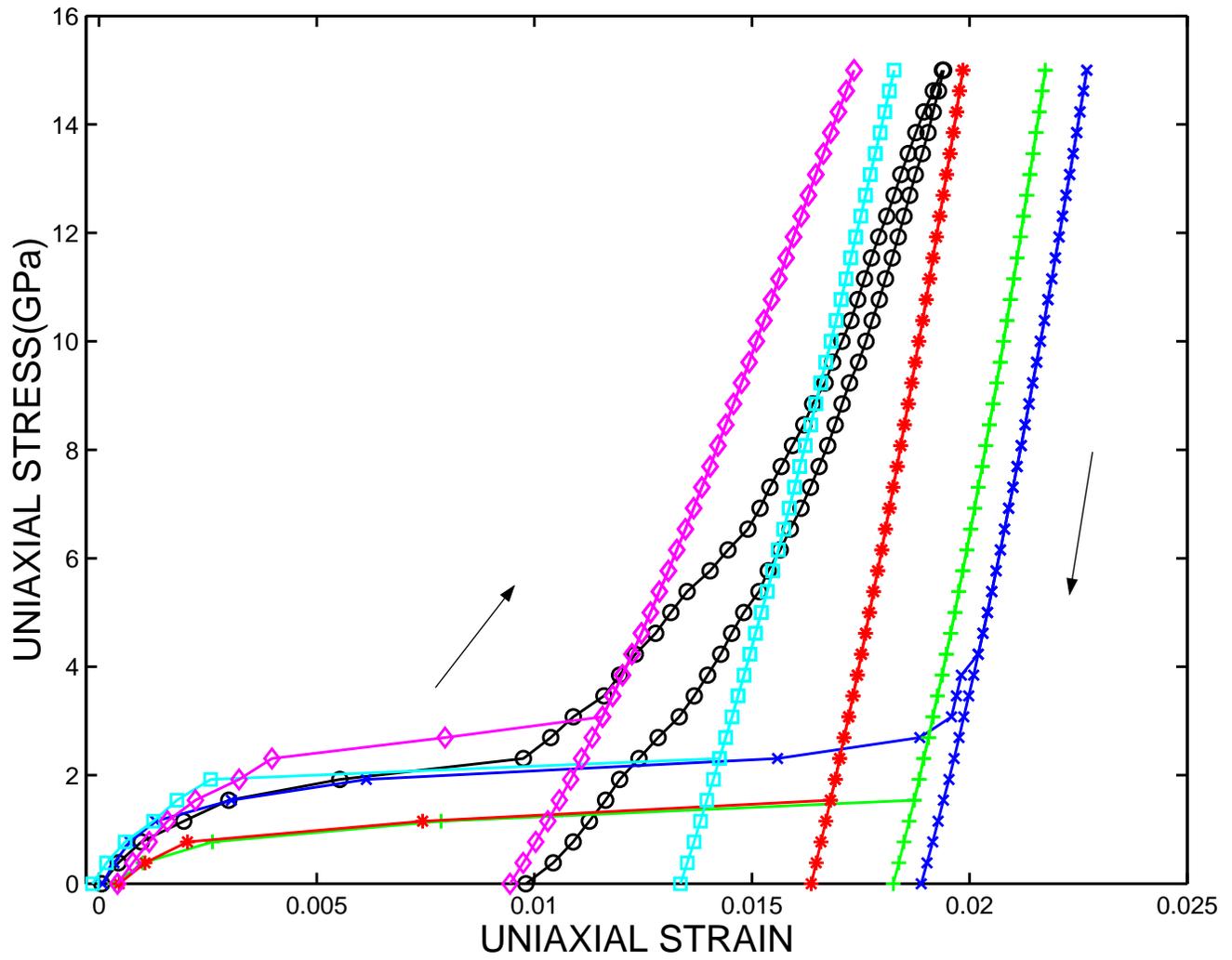
\end{document}